# Nonintegrability, chaos, and complexity

Joseph L. McCauley

Physics Department

University of Houston

Houston, Texas 77204

## Abstract

Two dimensional driven-dissipative flows are generally integrable via a conservation law that is singular at equilibria. Nonintegrable dynamical systems are confined to n≥3 dimensions. Even driven-dissipative deterministic dynamical systems that are critical, chaotic or complex have n-1 local time-independent conservation laws that can be used to simplify the geometric picture of the flow over as many consecutive time intervals as one likes. Those conservation laws generally have either branch cuts, phase singularities, or both. The consequence of the existence of singular conservation laws for experimental data analysis, and also for the search for scale-invariant critical states via uncontrolled approximations in deterministic dynamical systems, is discussed. Finally, the expectation of ubiquity of scaling laws and universality classes in dynamics is contrasted with the possibility that the most interesting dynamics in nature may be nonscaling, nonuniversal, and to some degree computationally complex.

## Introduction

It is expected that any system of ordinary differential equations generating critical (at a border of chaos), chaotic, or complex motions must be both nonlinear and "nonintegrable". The idea of "integrability" (complete integrability) vs. "nonintegrability" (or "incomplete integrability") goes back to Jacobi and Lie. The first example of a chaotic "nonintegrable" system was discovered via geometric analysis in phase space by Poincaré [1]. The first analytic evidence of "everywhere dense chaos" in a two-degree of freedom Hamiltonian system was discovered by Koopman and von Neumann [2]. Many experts still argue over the meaning of integrability vs. nonintegrability [3], and so misconceptions among physicists are abundant.

The ambiguity inherent in any serious attempt to distinguish "integrability" from "nonintegrability" was expressed poetically by Poincaré, who stated that a dynamical system is generally neither integrable nor nonintegrable, but is more or less integrable [4]. For physicists, the explanation of various roots to chaos (via period doubling [5], e.g.) has tended to submerge rather than clarify the question how to distinguish those two ideas, but without eliminating many of the misconceptions about nonintegrability. Modern mathematicians have managed to give precise definitions of nonintegrability [6b,7] that are very hard to translate into simpler mathematical language. With the explosion of interest in "complex dynamical systems" [8,9] it would be helpful to have a clearer idea, in ordinary words, of what "nonintegrability" means, especially as many articles about dynamics use terms that are either ambiguous or completely undefined: randomness, self-organization, self-organized criticality, complexity, complex adaptable system, and nonintegrability some are examples.

In an otherwise stimulating and revolutionary paper [9] Moore, has stated that for "... integrable systems ... a formula could be found for all time describing a system's future state." He also states that for chaotic systems "... we were forced to relax our definition of what constitutes a solution to the problem, since no formula exists. Instead, we content ourselves with measuring and describing the various statistical properties of a system, its scaling behavior and so on: we can do this because the individual trajectories are essentially random." These claims are often used to describe integrability and chaos at the level of popularized folklore [10], but they are wrong from both analytic and computational standpoints (Moore's article is about computability theory in the context of theoretical dynamics, not about experimental-data analysis). As a counter-example the binary Bernoulli shift, a so-called "paradigm" of "nonintegrability", has unique solutions that are given for all discrete times $n$ by the simple formula $x_n = 2^n x_0 \mod 1$. No "integrable" problem, including the simple harmonic oscillator, has a simpler solution (transcendental functions are required to describe the harmonic oscillator, whereas we only need simple algebra or easy decimal arithmetic to describe the Bernoulli shift). I will discuss "nonintegrability" and then argue that a misunderstanding of what it means can lead to uncontrolled approximations to critical, chaotic or complex dynamical systems that are both quantitatively *and* qualitatively wrong.

**Time reversibility and analyticity**

For the sake of clarity and precision I frame my discussion in the context of flows in phase space,

$$\frac{dx}{dt} = V(x), \qquad (1)$$

where phase space is a flat inner product space so that the n axes labeled by $(x_1,...,x_n)$ can be regarded as Cartesian [11], and $V(x)$ is an n-component time-independent velocity field. Newtonian dynamical systems can always be rewritten in this form whether or not the variables $x_i$ defining the system in physical three dimensional space are Cartesian (for example, we can have $x_1 = \theta$ and $x_2 = d\theta/dt$, where $\theta$ is an angular variable). Flows that preserve the Cartesian volume element $d\Omega = dx_1...dx_n$ are defined by $\nabla \cdot V = 0$ (conservative flows) while driven dissipative-flows correspond to $\nabla \cdot V \neq 0$, where $\nabla$ denotes the Cartesian gradient in n dimensions.

The condition for a flow is that for any initial condition $x_o$ the solution $x_i(t) = U(t)x_{io} = \psi_i(x_{o1},...x_{on},t)$ has no *finite* time singularities [6]; singularities of streamlines of flows are confined to the complex time plane. The time evolution operator $U(t)$ exists and defines a one parameter transformation group for all finite times t, with the inverse operator given by $U^{-1}(t) = U(-t)$, so that one can in principle integrate backward in time, $x_{oi} = U(-t)x_i(t) = \psi_i(x_1(t),...x_n(t),-t)$, as well as forward. Note that the same function $\psi_i$ describes integrations both forward and backward in time.

Many chaos researchers believe in floating point arithmetic the way that fundamentalists believe in heaven and hell (the latter cite the Bible for support, while the former rely on a literal interpretation of a purely formal abstraction called "the shadowing lemma"). Uncontrollable errors are introduced into numerical integrations by the use of floating point arithmetic and those errors violate time reversibility even in the simplest possible cases. Even for a *nonchaotic* driven-dissipative flow those errors prevent accurate numerical solutions either forward or backward in time [11]. The simplest example is given by the one dimensional flow $dy/dt = y$, all of whose streamlines have the positive Liapunov exponent $\lambda = 1$ forward in time, and the negative Liapunov exponent $\lambda = -1$ backward in time. Consequently, the simple linear equation $dy/dt = y$ cannot be integrated forward in time accurately numerically, for long times, if floating point arithmetic is used.

Contrary to superficial appearances based upon the unwarranted extrapolation of numerical calculations, time reversal is not violated by the Lorenz model

$$\frac{dx_1}{dt} = \sigma(x_2 \pm x_1)$$

$$\frac{dx_2}{dt} = \rho x_1 \pm x_2 \pm x_1 x_3$$

$$\frac{dx_3}{dt} = \pm \beta x_3 + x_1 x_2 \qquad . \quad (1b)$$

Let $z_n$ denote the maxima of a time series [12] of $x_3(t)$ at discrete times $t_o$, $t_1$, ..., $t_n$, ... , where $t_n - t_{n-1}$ denotes the time lag between successive maxima $z_{n-1} = x_3(t_{n-1})$ and $z_n = x_3(t_n)$. Formally, by paying attention to the implicit function theorem, it is possible to reduce any three dimensional flow to a uniquely-invertible two dimensional iterated map [11]

$$x_n = G_1(x_{n-1}, z_{n-1})$$
$$z_n = G_2(x_{n-1}, z_{n-1}).$$
$$(1c)$$

Uncontrolled numerical integrations suggest the one dimensional cusp map $z_n = f(z_{n-1})$ shown in figure 1. No one has yet explained why the Lorenz model should give rise, in any applicable approximation, to a *one* dimensional map. To the extent that the Lorenz model can be described approximately by a one dimensional map, that map can not have a double-valued inverse $z_{n-1} = f^{-1}(z_n)$: backward integration $z_{n-1} = U(t_{n-1} - t_n)z_n = \psi_3(x_n, y_n, z_n, t_{n-1} - t_n)$ is unique for a flow, and the Lorenz model satisfies the boundedness condition for a flow [12]. Therefore, Lorenz's one dimensional cusp map $z_n = f(z_{n-1})$ of figure 1 is not continuous and may even be infinitely fragmented and nondifferentiable in order that the inverse map $f^{-1}$ doesn't have two branches.

In the field of cosmology, the assumption that initial conditions of the early universe can be discovered from observable galaxy distributions can easily be challenged by anyone who has tried to integrate the double pendulum equations of motion forward for not very-long times, while using floating point arithmetic, and then backward in time in an attempt to recover even one digit of an initial condition [11].

Surprise has been expressed (perhaps ironically, in order to stress a point) that it was found possible to describe a certain chaotic flow by a formula in the form of an infinite series [3], but "nonintegrable" can not mean not solvable:

any flow, even a critical, chaotic or complex one, has a unique, well-defined solution so long as the velocity field V(x) satisfies a Lipshitz condition (or is at least once continuously differentiable) with respect to the n variables $x_i$. If, in addition, the velocity field is analytic in those variables then the power series

$$x_i(t) = x_{io} + t\,(Lx_i\,)_o + t^2\,(L^2x_i\,)_o\,/2 + \ldots, \quad (2)$$

where $L = V \cdot \nabla$, has a nonvanishing radius of convergence, so that the solution of (1) can in principle be described by power series combined with analytic continuation for *all* finite times [1]. It has long been known that this is not a practical prescription for the calculation of trajectories at long times. The point is that a large category of deterministic chaotic and perhaps even complex motions are precisely determined over any desired number of finite time intervals by *analytic formulae.* The Lorenz model (1b) provides an example. This is impossible for the case of truly "random" motion (like $\alpha$-particle decays), where the specification of an initial condition does not determine a trajectory at all, or for Langevin descriptions of diffusive motion, where the trajectories are continuous but are everywhere nondifferentiable (as in Wiener's functional integral). In uncontrolled approximations these separate ideas are sometimes confused together in a way that is impossible to untangle.

## Complete integrability

To define the idea of complete integrability we return to Jacobi [7] and Lie [13]. A completely integrable dynamical system has n-1 time-independent first integrals (conservation laws) $G_i(x_1,...,x_n) = C_i$ satisfying the linear partial differential equation

$$\frac{dG_i}{dt} = V \cdot \nabla G_i = V_k \frac{\partial G_i}{\partial x_k} = 0 \qquad (3)$$

along any streamline of the flow. In addition, these conservation laws must (in principle, but not necessarily constructively) determine n-1 "isolating integrals" of the form $x_k = g_k(x_n, C_1,...,C_{n-1})$ for k = 1,...,n-1. When all of this holds then the global flow is simply a time-translation for *all* finite times t in the Lie coordinate system

$$y_i = G_i(x_1,...,x_n) = C_i, \ i = 1,...,n\text{-}1$$
$$y_n = F(x_1,...,x_n) = t + D \qquad (4)$$

defined by the n-1 conservation laws, and the system is called completely integrable: the solution reduces *in principle* to n *independent* (rather than coupled, as in Picard's iterative procedure) integrations, and the flow is confined to a two-dimensional manifold that may be either flat or curved and is determined by the intersection of the n-1 global conservation laws (for a canonical Hamiltonian flow with f degrees of freedom, f commuting conservation laws confine the flow to a constant speed translation an f dimensional flat manifold). The $n^{th}$ transformation function $F(x_1,...,x_n)$ is defined by integrating $dt = dx_n/V_n(x_1,...,x_n) = dx/v_n(x_n,C_1,...,C_{n-1})$ to yield t + D = f($x_n$,$C_1$,...,$C_{n-1}$). One then uses the n-1 conservation laws to eliminate the constants $C_i$ in favor of the n-1 variables $x_i$ in f to obtain the function F. Whether one can carry out all or any of this constructively, in practice, is geometrically irrelevant: in the description (4) of the flow all effects of interactions have been eliminated globally via a coordinate transformation. The transformation (4) "parallelizes" (or "rectifies" [6]) the flow: the streamlines of (1) in the y-coordinate system are parallel to a single axis $y_n$ for all times, where the time evolution operator is a simple time-translation $U(t) = e^{td/dy_n}$.

Although time-dependent first integrals are stressed in discussions of integrable cases of driven-dissipative flows like the Lorenz model [3], there is generally no essential difference between (3) and the case of n time-dependent first integrals $G'_i(x_1,...,x_n,t) = C'_i$ satisfying

$$\frac{dG_i}{dt} = V \cdot \nabla G_i + \frac{\partial G_i}{\partial t} = 0 \qquad (3b)$$

Paying attention to the implicit function theorem, one conservation law $G'_n(x_1,...,x_n,t) = C'_n$ can be used to determine a function $t = F'(x_1,...,x_n,C'_n)$, whose substitution into the other n-1 time-dependent conservation laws yields n-1 time-independent ones satisfying (3). The n initial conditions $x_{io} = U(-t)x_i(t)$ of (1) satisfy (3b) and therefore qualify as time-dependent conservation laws, but initial conditions of (1) are generally only *trivial local* conservation laws: dynamically seen, there is no qualitative difference between backward and forward integration in time. *Nontrivial* global conservation laws are provided by the initial conditions $y_{io}$, for i = 1, 2, ... , n-1, of a completely integrable flow in the Lie coordinate system (4), where the streamlines are parallel for all finite times: $dy_i/dt = 0$, $i = 1,...,n-1$, and $dy_n/dt = 1$.

According to the Lie transformation (4) a completely integrable noncanonical flow is equivalent to a constant speed translation and is confined for all finite times t to a two-dimensional manifold. That manifold may be either curved or flat, and algebraic or at least analytic conservation laws [3] have generally been assumed to be necessary in order to obtain complete integrability. For example, Euler's description of a torque-free rigid body [11, 14]

$$\frac{dL_1}{dt} = a\, L_2 L_3$$

$$\frac{dL_2}{dt} = \pm b L_1 L_3$$

$$\frac{dL}{dt} = c\, L_1 L_2 \qquad , \quad (5)$$

with positive constants $a$, $b$, and $c$ satisfying $a - b + c = 0$, defines a phase flow in three dimensions that is confined to a two dimensional sphere that follows from angular momentum conservation $L_1{}^2 + L_2{}^2 + L_3{}^2 = L^2$. Here, we have completely integrable motion that technically violates the naive expectation that each term in (4) should be given by a *single* function: for each period $\tau$ of the motion, the transformation function F has four distinct branches due to the turning points of the three Cartesian components $L_i$ of angular momenta on the sphere. In general, any "isolating integral" $g_k$ describing bounded motion must be multivalued at a turning point.

Since the time of Kowalevskaya some mathematicians have defined first integrals $G_k$ of dynamical systems to be analytic or at least continuous [6] (however, see also ref. [6b] where nonanalytic functions as first integrals are also discussed). This is an arbitrary restriction that is not always necessary in order to generate the transformation (4) over all finite times: a two-dimensional flow in phase space, including a driven-dissipative flow, is generally integrable via a conservation law but that conservation law is typically singular. The conservation law is simply the function that describes the two-dimensional phase portrait and is singular at sources and sinks like attractors and repellers (equilibria and limit cycles provide examples of attractors and repellers in driven-dissipative planar flows) [11]. For the damped simple harmonic oscillator, for example, the conservation law has been constructed analytically [15] and is logarithmically singular at the sink. The planar flow where $dr/dt = r$ and $d\theta/dt = 0$ in cylindrical coordinates $(r,\theta)$ describes radial flow out of a source at $r = 0$. The conservation law is simply $\theta$, which is constant along every streamline and is undefined at $r = 0$. This integrable flow is parallelizeable for all *finite* times t simply by excluding one point, the source at $r = 0$ (infinite time would be required to leave or reach an equilibrium point, and no one should care what happens in the completely unphysical limit of infinite time). "Nonintegrable" flows generally can not occur in the phase plane. What about in three or more dimensions?

**"Nonintegrability"**

A completely integrable flow is described by the formulae (4) for all finite times t and is parallel to a single axis $y_n$ (the n Lie coordinates $y_i$ are generally locally orthogonal on a flat or curved two dimensional manifold). In contrast, there is also an idea of *local* integrability [6,16]: one can parallelize an arbitrary flow (including chaotic and complex ones) about any nonequilibrium point, meaning about any point $x_o$ where the velocity field $V(x)$ does not vanish. The size $\varepsilon(x_o)$ of the region where this parallelization holds is *finite* and depends nonuniversally on the n gradients of the velocity field. By analytic continuation [6b,17], local parallelization of the flow yields n-1 nontrival "local" conservation laws $y_i = G_i(x) = C_i$ that hold out to the first singularity of any one of the n-1 functions $G_i$, in agreement with the demands of the theory of first order linear partial differential equations (the linear partial differential equation (3) always has n-1 functionally independent solutions, but the solutions may be singular [15]).

Consider the streamline of a "nonintegrable" flow that passes through a nonequilibrium point $x_o$ at time t = 0. Let $t(x_o)$ then denote the time required for the trajectory to reach the first singularity of one of the conservation laws $G_k$. Such a singularity must exist, otherwise the flow would be confined for *all* finite times ("globally") to a single, smooth two-dimensional manifold. The global existence of a two-dimensional manifold can be prevented, for example, by singularities that make the n-1 conservation laws $G_i$ multivalued in an extension of phase space to complex variables [6b]. Generally, as with the solutions defined locally by the series expansion (2), the n-1 local conservation laws $G_i$ will be defined locally by infinite series, with radii of convergence determined by the singularities in the complex extension of phase space. The formulae (4) then hold for a finite time $0 \leq t < t(x_o)$ that is determined by the distance from $x_o$ to the nearest complex singularity. Let $x_1(x_o)$ denote the point in phase space where that singularity occurs. Following Arnol'd's [6] statement of the "basic theorem of ordinary differential equations", we observe that the streamline of a flow (1) passing through $x_o$ can not be affected by the singularity at $x_1$ in the following superficial sense (consistent with the fact that the singularities of the functions $G_i$ are either branch cuts or phase singularities): we can *again* parallelize the flow about the singular point $x_1(x_0)$ and can *again* describe the streamline for *another* finite time $t(x_0) \leq t < t(x_1)$ by another set of parallelized flow equations of the form (4), where $t(x_1)$ is the time required to reach the next singularity $x_2(x_0)$ of any one of the n-1 conservation laws $G_i$, starting from the second initial condition $x_1$. Reparallelizing the flow about any one of these singularities reminds us superficially of resetting the calendar when crossing the international dateline, excepting that a nonintegrable flow is not confined to a globally analytic two dimensional manifold. Two points follow.

First, a "nonintegrable" flow is "piecewise integrable" in the sense that different sets of formulae of the form (4) hold *in principle* for consecutive finite time intervals $0 \leq t(x_0) < t(x_1)$, $t(x_1) \leq t < t(x_2)$, ... $t(x_{n-1}) \leq t < t(x_n)$, .... , giving geometric meaning to Poincaré's dictum [3] that a dynamical system is generally neither integrable nor nonintegrable but is more or less integrable. Nonintegrable flows are describable over arbitrarily-many consecutive time intervals by the simple formulae of the form (4) except at countably many singular points $x_1(x)$, $x_2(x)$, ... , where the n-1 initial conditions $y_{io}$ and the constant D must be reset. Contrast this with the claim of ref. [9] discussed above where the use of the word "formula" was not restricted (there are *completely integrable* cases where no formula for F in (4) has been "found", if the word "found" is construed to mean *construction* rather than merely mathematical *existence*). The distinction between completely and incompletely integrable dynamical systems is therefore technical and difficult to make, but it is not purely semantic (see chapter 6 of reference [6b] for a more detailed discussion of the causes of "nonintegrability").

Second, we consider the implication for interpretations of Taken's embedding theorem [18,19], which has sometimes led to the expectation that the dimension n of a chaotic dynamical system may be computationally compressible (replacement of a high dimensional flow by a lower dimensional one). Consider a critical or chaotic flow (1) with dimension n≥4. If we should find that some class of trajectories (meaning some class of initial conditions [11]) yields a fractal dimension $D_o \approx 2.1$, e.g, then this would not mean that the flow can be studied for long times on a *flat* three dimensional manifold: a three dimensional system of differential equations describing a flow (1) with n≥4 could only arise from the elimination of n-3 variables via n-3 generally singular conservation laws, which would place the flow on a section of a three dimensional manifold that is generally both singular and *curved*, and even then only over a finite time interval bounded by the time $t(x_o)$ out to the first singularity starting from some initial condition $x_o$. In other words, that a flow may be locally three dimensional does not mean that we can lift it out of ten dimensions and embed it globally in three dimensions. Smooth m dimensional manifolds that can be embedded in n dimensions are not necessarily embeddable in m<n dimensions: four dimensional geodesic flows on compact two dimensional manifolds of constant negative curvature provide an example [20].

*A topologically-correct description of a higher dimensional chaotic dynamical system by a lower dimensional one is possible, via symbol sequences and their statistical distributions, but only if the two systems belong to the same topologic universality class [21,21b] (spectra of Liapunov exponents and fractal dimensions are not topologic invariants).*

We argue next, heuristically, that the existence of singular conservation laws for deterministic dynamical systems has important consequences for approximations and modelling in theoretical physics.

**Universality classes, statistics, and scaling laws**

There are many different phenomena in nature that seem, more or less, approximately to obey scaling laws of one sort or another. Since the success of the renormalization group method in describing approximate scaling laws near second order phase transitions in statistical mechanics, and since fractals have been found to occur in both critical and chaotic deterministic dynamical systems, many physicists have come to believe that scaling laws may be ubiquitous in nature. Some physicists even expect a general explanation of scaling laws that occur in nature in terms of universally-valid dynamics that yield scaling exponents free of parameter-tuning.

One prescription for scale-invariant correlations is supposed to be a driven-dissipative system with one conserved quantity [22], with the dynamics perturbed externally by "random noise". A partial realization of this prescription has been demonstrated for a certain class of linear partial differential equations satisfying special boundary conditions. The addition of nonlinear terms breaks the asymptotic scale invariance reflected in the correlation function $G(r) \approx r^{-d}$ of the linear theory, where d is the dimension of the system (compare with $G(r) \approx r^{-d+2}$ for the Laplace equation when d>2) if those terms are significant enough not to be perturbatively approximable in the renormalization group method, which is usually an uncontrolled approximation. A related question is whether an intractable nonlinear equation can be replaced by a more tractable one subject to external noise [22], as in comparisons of the deterministic Kuramoto-Sivashinsky partial differential equation with the noise-driven KPZ equation [23]. We expect that the conclusions stated below for flows will carry over to diffusive deterministic partial differential equations.

In a deterministic dynamical system (1) that is either critical or chaotic, a qualitatively-correct description of the motion via a perturbative treatment of nonlinear terms in an otherwise linear (or nonlinear but integrable) system is impossible. In particular, neither chaotic nor critical orbits can be described by noisy orbits of an otherwise linear (or nonlinear but integrable) system. The replacement of intractable nonlinear terms by "random noise" in an uncontrolled approximation may violate geometric constraints imposed by the deterministic system's local conservation laws.

In contrast with reasoning based upon nonlinear or random perturbations of linear dynamics, the addition of linear damping and driving to an important class of deterministic conservative nonlinear systems (nonstandard Euler-Lagrange equations with nonintegrable velocities like angular momenta as variables, like (5) [11], e.g.) leads to self-confined motion and, via bifurcations, to nontrivial critical behavior and chaotic attractors in phase space [12]. The Lorenz model provides a simple example. Formally, the Lorenz model defines a certain linearly damped and driven symmetric top: simply set a = 0, and b = c = 1, in (5), which then coincides with (1b) if we throw away all of the *linear* terms in the latter.

The prescription for noise-driven scale-invariant correlations is based on the expectation that conservation laws are not typical for driven-dissipative dynamical systems, that by imposing "local conservation" [22] we can change the dynamics. Because dynamical systems (1) already have a complete set of local first integrals the freedom of choice that would permit the imposition of extra constraints does not exist in the case of deterministic ordinary differential equations.

A refinement of the condition of "local conservation" has been offered as a definition of "self-organized criticality" (SOC) by Anderson [8], who states that "a system driven by some conserved or quasi-conserved quantity uniformly at a large scale, but able to dissipate it only to microscopic fluctuations, may have fluctuations at all intermediate scales,... . The canonical case of SOC is turbulence ... ." See also reference [23] for a slightly more refined definition of SOC where the emphasis is on two widely-separated time scales, a very short time scale for slow external driving at large length scales vs. a relatively long time scale for dissipation via diffusion at very short length scales (as is characteristic of eddy-cascades at all Reynolds numbers in fluid dynamics).

Anderson's attempt to define SOC would describe fluid turbulence in open flows (the Richardson-Kolmogorov eddy-cascade at high Reynold's numbers, e.g.) if we could replace the word "fluctuations" with the phrase "a hierarchy of eddies where the eddy-cascade is generated by successive instabilities". The empirical evidence suggests that neither laminar nor turbulent eddy-cascades are critical (meaning *metastable* behavior [23] dominated by a vanishing Liapunov exponent): vortex-instability cascades yield rapid mixing even at low Reynolds numbers [24] of 15 to 20, and rapid mixing suggests the action of at least one positive Liapunov exponent. Correspondingly, the dissipation rate per unit mass for the dissipation range of turbulent open flows [24] suggests a Liapunov exponent on the order of ln2.

From an entirely different and very formal perspective it has been argued that it may be possible to replace the Navier-Stokes equations asymptotically by a finite dimensional autonomous phase flow of the form (1) [25,26], but it is not clear whether this argument can be correct.

No model exhibiting both criticality and scaling laws that are independent of parameter tuning has ever been constructed from any set of deterministic differential equations, including block-spring models, which are supposed to represent the essence of SOC [22,27]. In the models used to try to define SOC, universality classes are supposed to be identified on the basis of one or a few scaling exponents, but a few scaling exponents are generally not adequate to define a dynamical universality class unambiguously in the interaction of many degrees of freedom far from thermal equilibrium.

I explain my assertion via an example that is apparently unstable rather than metastable: in the inertial range of fluid turbulence the experimental data are precise enough to pin down only one velocity structure function scaling exponent $\zeta_o$, which in turn determines only one fractal dimension $D_o = 3 - \zeta_o \approx 2.91$. Infinitely many different multifractal generating functions are consistent with the experimentally-measured *nonlinear* scaling exponents $\zeta_p$ for p>10 [24,28]. Each multifractal generating function, in turn, is consistent, to within experimental accuracy, with infinitely many different and completely unrelated abstract (and generally unphysical) dynamical models, some deterministic and some stochastic. My point is simple: it is too easy to construct an ad hoc model with no known connection to the Navier-Stokes equations that reproduces the *entire spectrum* of velocity structure function scaling exponents *to within experimental accuracy*.

A sharper result is known for the measured energy transfer rate near the Kolmogorov length scale: the experimentally-extracted statistics (binomial, with $p_1/p_2 \approx 3/7$) and generating partition ($l_n \approx 2^{-n}$) are trivially reproduced by the binary tent map for a "measure zero" class of initial conditions. The binary tent map has no known connection to the physics of eddy cascades, other than that complete binary-cascading is observed visually near the Kolmogorov limit of the dissipation range, for a local Reynolds number near unity, in the the ink-droplet experiment [28]. In the inertial range the vortex birth rate suggested by the fit of the β-model, in the determination of $\zeta_o$, is on the order of octal and incomplete. For nonturbulent cascades the observed birth rate ranges from (incomplete) pentagonal or hexagonal at a local Reynolds number of 15 down to complete binary for R ≈ 1. For more recent discussions of scaling law phenomenology for eddy cascades in the inertial and dissipation ranges, see references [28b,c, and d].

A few scaling exponents can be used to distinguish different classes of one dimensional unimodal maps from each other at the period doubling critical point, but in this case the order of the maximum of a unimodal map (via a nonperturbative renormalization group method) defines a universality class analogous to the universality classes defined by symmetry and dimension at second order phase transitions in equilibrium statistical mechanics [29]. In the case of SOC, no corresponding definition has been given, which is essentially the same as saying that a clear definition of SOC has not yet been given. In the case of so-called "complex adaptable systems" [8], there is no attempt to universality classes. Hence, there is effectively no definition of "complex adaptable systems". It is not clear that a sand pile automaton, or an arbitrary model of a so-called "complex adaptable system", can provide a topologically-correct description of any dynamical system that occurs in physics, chemistry, or biology. In the case of sand pile models, some block spring models agree in part with SOC ideas (but for a system that is chaotic rather than critical [27b]), whereas others do not [27].

**Complexity vs. scaling and universality**

Scale invariance based upon criticality has been offered as an approach to "complex space-time phenomena" based upon the largely unfulfilled expectation of finding that universal scaling laws, generated dynamically by many interacting degrees of freedom and yielding critical states that are independent of parameter-tuning [22,23], are ubiquitous in nature. This is equivalent to expecting that nature is mathematically relatively simple, something that Newton, to his severe dismay, learned to be false when he invented perturbation theory in an attempt to solve the three-body problem for the moon's motion. Many physicists apparently can not understand why Newton gave up science for alchemy, religion, politics and finance, but then most physicists have never attempted to solve the three body problem.

The incompressible Navier-Stokes equations are formally invariant under the scaling transformation $x' = \lambda x$, $v' = \lambda^{\alpha} v$, and $t' = \lambda^{1-\alpha} t$, which leaves the Reynolds number invariant. This scale invariance encourages the expectation that the velocity structure functions may be scale invariant with exponents $\zeta_p$ as well, and also permits the formulation of the idea that the $f(\alpha)$ spectrum may be useful for describing a distribution of velocity-field singularities that may occur in the limit of infinite Reynolds numbers. With due respect for the program of research that is advocated in ref. [22,23], however, the greater problem for science may be to discover and account for turbulent flows where the velocity structure functions do not obey scaling laws. To discover nonscaling behavior is probably not difficult experimentally, but under the expectation that the most significant phenomena in nature obey scaling laws and some sort of universality such results would likely be classified by many physicists as "uninteresting".

It has recently been discovered that there is a far greater and far more interesting degree of complicated behavior in nonlinear dynamics than either criticality or deterministic chaos: systems of billiard balls combined with mirrors [30], and even two-dimensional maps [9,31] can exhibit universal computational capacity (via formal equivalence to a Turing machine). A system of nine first order quasi-linear partial differential equations has been offered as a computationally-universal system [32]. A quasi-linear first order partial differential equation in n variables can be replaced by a linear one in n+1 variables. *Maximum computational complexity is apparently possible in systems of linear first order partial differential equations*. When is complexity of this sort *not* allowed?

If a deterministic dynamical system has a generating partition [21] then the symbolic dynamics can in principle be solved and the future behavior can be understood qualitatively, without the need to compute specific trajectories algorithmically from the algorithmic construction of a specific computable initial condition [24]. In other words, a high degree of "computational compressibility" holds even if the dynamical system is orbitally-metastable ("critical") or orbitally-unstable ("chaotic").

A chaotic dynamical system generates infinitely-many different classes of statistical distributions for different classes of initial conditions (at most one distribution is differentiable). The generating partition, if it exists, uniquely forms the support of every possible statistical distribution and also characterizes the particular dynamical system. For phase flows and iterated maps, criticality and deterministic chaos can be distinguished from dynamics with universal computational complexity by the existence of a generating partition. For a system with a generating partition, topologic universality classes exist that permit one to study the simplest system in the universality class (the infinity of statistical distributions is topologically invariant and therefore can *not* be used to discern or characterize a particular dynamical system in a universality class). For maps of the unit interval, both the symmetric and asymmetric logistic maps of the unit interval peaking at or above unity belong to the trivial universality class of the binary tent map [24]. The two dimensional Henon map is much harder but is still solvable: it belongs to the universality class of chaotic logistic maps of the unit interval peaking beneath unity [21] (the simplest model defining this universality class is the symmetric tent map with slope magnitude between 1 and 2). In these systems the long-time behavior can be understood qualitatively and statistically *in advance*, so that the future holds no surprises: the generating partition and symbol sequences can be used to describe the motion at long times to within any desired degree of precision, and multifractal scaling laws (via the $D(\lambda)$ spectrum, e.g.) show how finer-grained pictures of trajectories are related to coarser-grained ones.

For a dynamical system with universal computational capability, in contrast, a classification into topologic universality classes is impossible [9]. Given an algorithm for the computation of an initial condition to as many digits as computer time allows, nothing can be said in advance about the future *either statistically or otherwise* other than to compute the dynamics, with controlled precision for that initial condition, iteration by iteration, to see what falls out: there is no computational compressibility that allows us to summarize the system's long-time behavior, either statistically or otherwise. In contrast with the case where topologic universality classes exist there is no organization of a hierarchy of periodic orbits, stable, marginally stable, or unstable, that allows us to understand the fine-grained behavior of an orbit from the coarse-grained behavior via scaling laws, or to divine the very distant future for arbitrary (so-called "random") initial conditions from the symbolic dynamics. Certainly, there can be no scaling laws that hold independently of a very careful choice of classes of initial conditions, if at all. We do not know if fluid turbulence or Newton's three-body problem fall into this category.

## Social Darwinism and Malthusian biology

Contrary to recent expectations [18,34] and extraordinary claims[a], there is no evidence to suggest that abstract dynamical systems theory[b] can be used either to explain or understand socio-economic behavior. Billiard balls and gravitating bodies have no choice but to follow mathematical trajectories that are laid out deterministically, *beyond the possibility of human convention, invention, or intervention*, by Newton's laws of motion. The law of probability of a Brownian particle evolves deterministically according to the diffusion equation, also *beyond the possibility of human convention, invention, or intervention.* In stark contrast, a mind that directs the movements of a body continually makes willful and arbitrary decisions at arbitrary times that cause it to deviate from any mathematical trajectory (deterministic models) or evolving set of probabilities (stochastic models) assigned to it in advance. Given a hypothetical set of probabilities for a decision at one instant, there is no algorithm that tells us how to compute the probabilities correctly for later times (excepting at best the trivial case of curve-fitting at very short times, and then only if nothing changes significantly), or even when later decisions will be made. Two disputants in an argument, or two promenaders on a path, can think and choose to alter their courses in order to avoid a collision (or can choose to collide), whereas two billiard balls, or two planets, on a collision course have no similar possibility.

Socio-economic statistics can not be known in advance of their occurrence because, to begin with, there are no known, correct socio-economic laws of

---

motion. We can describe and *understand* tornadoes and hurricanes mathematically because Newton's laws apply, in spite of the fact that the earth's atmosphere is an *open* dynamical system, but we can not *understand* the collapse of the Soviet Union or the financial crisis in Mexico on the basis of any known set of dynamics equations, in spite of the fact that the world economy forms a *closed* financial system. I have argued elsewhere that trajectories and probabilities for socio-economic "motion", which are only motions in an Aristotelian rather than Galilean sense, are not computable [35].

Some degrees of complexity are defined precisely in computer science [36] but these definitions have not satisfied some physicists [8,37,37b]. According to von Neumann [38] a system is complex when it is easier to build than to describe mathematically. Under this qualitative definition the Henon map is not complex but a living cell is. Earlier claims to the contrary [39,40], where complexity was confused with information, there is as yet no model of a dynamic theory of the *evolution* of biologic complexity, not over short time intervals (cell to embryo to adult) and certainly not over very long time intervals (inorganic matter to organic matter to metabolizing cells and beyond). Correspondingly, there is no physico-chemical definition of the development of different degrees of complexity in nonlinear dynamics. No one knows if universal computational capability is necessary for biologic evolution, although DNA molecules in solution apparently can be made to compute [41] (but not error-free, as with a Turing machine or other deterministic dynamical system).

It has been speculated that computational universality should be possible in a conservative three degree of freedom Newtonian model [9], but so far no one has constructed an example. We do not yet know the minimum number of degrees of freedom necessary for universal computational capability in a set of driven-dissipative equations (1) (the Omohundro model [32] is driven-dissipative). Diffusive motion is generally time-irreversible ($U^{-1}(t)$ doesn't exist for diffusive motion), but arguments have been made that some diffusive dynamical systems have an asymptotic limit where the motion is time reversible on a finite dimensional attractor [26.42], and is therefore generated by a finite dimensional deterministic dynamical system (1). However, if a diffusive dynamical system (the Navier-Stokes equations, e.g.) should be found to be computationally-universal then it will be impossible to discover a single attractor that would permit the derivation of scaling laws for eddy cascades in open flows, or in other flows, *independently of specific classes of boundary and initial conditions.*

Whenever universal computational capability is generated by a deterministic or diffusive dynamical system then the usual notions of universality classes in statistical physics and deterministic chaos fail, and that dynamical system can only be studied in special cases defined by specific classes of boundary and initial conditions. It seems unlikely that scaling and universality classes will be useful in the understanding of biology, which is too hard to be left entirely to the biologists [43] [c].

---

# Reductionism and Newtonian complexity

"Reductionism" is criticized by "holists" (see the introduction to ref. [8] and also [47], e.g.). Some holists propose to mathematize Darwinism in order to go beyond physics and chemistry, but they have not been able use dynamics models to predict or explain anything that occurs in nature. Physics and astronomy, since the divorce from Platonic mathematics and Aristotlelian "holism" in the seventeenth century, have a completely different history (or "evolution") than "political economy" and most of biology.

Reductionism is the arbitrary division of nature into laws of motion and initial conditions, plus "the environment". We must always be able to neglect "the environment" to zeroth order, because if *nothing* can be isolated then a correct law of motion can never be discovered. Successful reductionism stems from Galileo and Descartes, whose discoveries were essential for Newton's formulation of universally-valid laws of motion and gravity.

Motivated by the question why mathematics works *at all* in physics, astronomy, and chemistry, Wigner [48] has made two extremely important observations: (1) the empirical discovery of mathematical laws of motion that correctly describe nature is impossible in the absence of empirically-significant invariance principles, but (2) there are no laws of nature that can tell us the initial conditions. Following Wigner, laws of nature themselves obey laws called invariance principles, while initial conditions are completely lawless.

Lawlessness reigns supreme in the socio-economic fields, where nothing of any significance is left even approximately invariant by socio-economic evolution. This is the reason that artificial law ("law") is passed in legislatures in an attempt to regulate behavior. From the standpoint of nonlinear dynamics everything of socio-economic significance changes completely uncontrollably, so that there are not even any fundamental constants.

About initial conditions that occur in nature, theoretical physics, and especially pure mathematics, has nothing at all to say. Initial conditions matter when a dynamical system far from thermal equilibrium is chaotic or complex. The only way to know an "initial condition" (or "present condition" as Lorenz [12] says) or distribution of initial conditions *approximately* is to consult mother nature via measurement. In a driven-dissipative system far from equilibrium, it is an illusion to expect that we replace our ignorance of initial conditions by the assumption of "random external noise" like "white noise" or Gaussian statistics. For example, no one anticipated that the dissipation range of turbulent open flows would generate binomial statistics with *uneven* probabilities $p_1/p_2 = 3/7$. There is still no physical explanation of this simple result.

There is certainly no observational evidence to indicate that evolving nature respects the abstractions that pure mathematicians try to sell to us under the heading of "invariant densities" and "random initial conditions". Assuming that the motion of a dynamical system is adequately described by assuming "random initial conditions" is the same as assuming that a deterministic system evolved from earlier initial conditions into a later state where the coarsegrained statistics are even: whether or not this assumption is true or false for a given system observed in nature can be decided only empirically by measurement, and not by pure mathematics or computer simulations.

The notion that nature far from equilibrium can described by "random initial conditions" is merely a *belief* that is not supported by any known set of measurements of nature. In computation, to assume "random initial conditions" is to presume that initial conditions are lawful: the law in question is simply the numerical algorithm used by the computer to construct a pseudorandom sequence of digits.

The best that can be done theoretically is to study the behavior of dynamical systems far from equilibrium for different classes of initial and boundary conditions without prejudicing those studies by assumptions for which there is no experimental or observational evidence. That approach is implicit in the method of classification of deterministic dynamical systems into topologic universality classes.

From the standpoint of Wigner's simple and very beautiful considerations, I argue that the belief that the initial conditions of the early universe can be discovered on the basis of physical principles alone (perhaps combined with illegal backward integrations that use floating point arithmetic) is an illusion. Any time that a physics paper appears to answer questions that are asked by theology, then the physicist who does the writing is likely under the influence of one illusion or another.

With a computationally-universal (and therefore computable) dynamical system (1), given a specific computable initial condition $x_o$, both that initial condition and the dynamics can in principle be encoded as the digit string for another computable initial condition $y_o$. The computable trajectory $y(t) = U(t)y_0$ might then in principle be digitally decodable, which would allow us to learn the trajectory $x(t) = U(t)x_0$ for the first initial condition (self-replication without copying errors). *This maximum degree of computational complexity may be possible in low dimensional nonintegrable conservative Newtonian dynamics* [9,27]. Some features of nonintegrable quantum systems with a chaotic classical limit (the helium atom, e.g.) have been studied using uncontrolled approximations based on the low order unstable periodic orbits of a chaotic dynamical system [49], but we have no hint what might be the behavior of a low dimensional quantum mechanical system with a computationally-complex Newtonian limit.

Physics has been declared to be dead from the standpoint of particle physics [50]. Correspondingly, some practitioners of quantum field theory (so-called "grand reductionists" [51][d]) have retreated from Galilean principles into Platonic mathematical speculations about cosmology that can never be tested adequately empirically [50].

From the perspective of Newtonian mechanics [9, 21b] and nonrelativistic quantum mechanics [49] (so-called "petty reductionism" [51]) it appears that the most interesting and significant work for physics has only barely begun.

### Acknowledgement

I am extremely grateful to my friend Julian Palmore for teaching me that a deterministic dynamical system (1) always has n-1 *nontrivial* time-independent conservation laws, and to Phil Morrison and Gemunu

---

d

Gunaratne for criticizing my earlier attempt to distinguish integrability from nonintegrability. I'm grateful to George Reiter for reading and criticizing on an earlier version of the manuscript. I'm also grateful to Geoff Grinstein and Joachim Krug for very clear and stimulating lectures on scale invariance, SOC and the Kuramoto-Sivashinsky and KPZ equations at the 1996 Brasilia Workshop on Complexity, to Charles Bennett for pointing out reference [9] in a Workshop lecture, to David Mukhamel for explaining to me the idea of trying to replace nonlinear terms by external noise in renormalization group arguments, and to Fernando de Oliveira for hospitality in Brasilia. Thanks are in order to to Stefan Metens for bringing references [25, 26, 42] to my attention at the *März 1996 Winterseminar auf dem Zeinisjoch*, and especially to Peter Plath for the invitation to speak there. Finally, I am also grateful to Per Bak for pointing out reference [27b], and for a telephone conversation that helped me to see and then eliminate the imprecision in an earlier version of this article.

### References

1. H. Poincaré, <u>New Methods of Celestial Mechanics</u> (AIP, Woodbury, NY, 1993).

2. B.O. Koopman and J. von Neumann, Proc. Nat. Acad. of Sc. <u>18</u> (1932) 255-263.

3. M. Tabor et al in <u>What is Integrability?</u>, ed. by V. E. Zakharov (Springer-Verlag, Berlin, 1991).

4. A. Wintner, <u>The Analytical Foundations of Celestial Mechanics</u>, sect. 194-202 & 227-240 (Princeton, Princeton, 1941).

5. M.J. Feigenbaum,, Los Alamos Science (1980).

6. V.I. Arnol'd,  Ordinary Differential Equations (M.I.T. Press, Cambridge, Mass., 1981).

6b. V. I. Arnol'd, V. V. Kozlov,  and A. I. Neishtadt, Mathematical Aspects of Classical and Celestial Mechanics,  in Dynamical Systems III, ed. by V.I. Arnold (Springer-Verlag , Heidelberg , 1993).

7. A.T. Fomenko,  Integrability and Nonintegrability in Geometry and Mechanics , transl. from Russian by M. V. Tsaplina (Kluwer, Dordrecht, 1988).

8.  G. A. Cowan, D. Pines, and D. Meltzer, editiors, Complexity, Metaphors, Models, and Reality (Addison-Wesley, Reading, 1994).

9. C. Moore, Phys. Rev. Lett. 64  (1990) 2354.

10. J. Gleick, Chaos (Viking, New York, 1987).

11. J.L. McCauley, Classical Mechanics: flows, transformations,  integrability, and chaos (Cambridge Univ. Pr., Cambridge, forthcoming in 1996).

12. E. Lorenz, J. Atm. Sc. 20 (1963) 130.

13. E.T. Bell, The Development  of Mathematics (McGraw-Hill, New York, 1945).

14. C.M. Bender and S.A. Orszag, Advanced Mathematical Methods for Scientists and Engineers , ch. 4. (McGraw-Hill, New York, 1978).

15. S.A. Burns and J.I. Palmore, Physica D37 (1989) 83.

16. P.J. Olver,  Applications of Lie Groups to Differential Equations , pg. 30 (Springer-Verlag, New York, 1993).

17. J. Palmore, private conversation (1995).

18. D. Ruelle, Physics Today (July 1994) pp. 24-30.

19. H. J. Schuster, <u>Deterministic Chaos</u>, 2nd revised ed. (VCH Verlagsgesellschaft, Weinheim, 1988).

20. V. I. Arnol'd, <u>Mathematical Methods of Classical Mechanics</u> (Springer-Verlag, Berlin, 1978).

21. P. Cvitanovic, G. Gunaratne, and I. Procaccia, Phys. Rev. A<u>38</u> (1988) 1503.

21b. G. H. Gunaratne in <u>Universality beyond the onset of chaos</u>, in <u>Chaos: Soviet and American Perspectives on Nonlinear Science</u>, ed. D. Campbell (AIP, New York, 1990).

22. G. Grinstein in <u>Scale Invariance, Interfaces, and Non-Equilibrium Dynamics</u>, ed. A. Mckane et al (Plenum, New York, 1995).

23. J. Krug, in <u>Modern Quantum Field Theory II</u>, ed. S. R. Das et al (World, Singapore, 1994).

24. J.L. McCauley, <u>Chaos, Dynamics, and Fractals: *an algorithmic approach to deterministic chaos*</u> (Cambridge Nonlinear Science Series 2, Cambridge, 1993).

25. P. Constantin, C. Foias, O. P. Manley, and R. Temam, J. Fluid Mech. <u>150</u> (1985) 427.

26. C. Foias, G. R. Sell, and R. Temam, J. Diff. Eqns. <u>73</u> (1988) 309.

27. A. Crisanti, M. H. Jensen, A. Vulpiani, and G. Paladin, Phys. Rev. <u>A46</u> (1992) R73653.

27b. M. de Sousa Viera and A. J. Lichtenberg, Phys. Rev. <u>E53</u> (1996) 1441.

28. J. L. McCauley in <u>Spontaneous Formation of Space-Time Structures and Criticality</u>, ed. D. Sherrington and T. Riste (Kluwer, Dordrecht, 1991).

28b. D. Lofse and S. Grossmann, Physica <u>A194</u> (1993) 519.

28c. R. Benzi, S. Ciliberto, R. Tripiccione, C. Baudet, F. Massaioli, and S. Succi (finally!), Phys. Rev. <u>E48</u> (1993) R29.

28d. Z.- S. She and E. Leveque, Phys. Rev. Lett. <u>72</u> (1994) 336.

29. B. Hu, Phys. Reports <u>31</u> (1982) 233.

31. E. Fredkin and T. Toffoli, Int. J. Theor. Phys. <u>21</u> (1982) 219.

31. C. Moore, Nonlinearity <u>4</u>  (1991) 199 &727.

32.  S. Omohundro, Physica <u>10D</u>  (1984) 128.

33. P. Bak and M. Paczuski, <u>Complexity, Contingency, and Criticality</u>, subm.  to Proc. Nat. Acad. Sci.  (1995).

34. P. W. Anderson,  K. J. Arrow, and D. Pines, eds., <u>The Economy as an Evolving Complex System</u> (Addison-Wesley, Redwood City, 1988).

35. J. L. McCauley, <u>Simulations,  complexity, and laws of nature</u>, accepted by Complexity (Feb., 1996).

36. M. R. Garey and D.S. Johnson,  <u>Computers and Intractability: A Guide to the Theory of NP-Completeness</u>  (Freeman, San Francisco, 1979).

37. C. Bennett, Found. Phys. <u>16</u> (1986) 585.

37b. S. Lloyd and H. Pagels, Annals of Phys. <u>188</u>, 186 (1988).

38. J. von Neumann, <u>Theory of Self-Reproducing  Automata</u>, pp. 47, 48 (Univ. of  Ill., Urbana, 1966).

39. M. Eigen, <u>Steps toward Life</u> (Oxford, Oxford, 1992).

40. B. -O. Küppers, <u>The Molecular  theory of Evolution</u>  (Springer,  Heidelberg, 1985).

41. Lipton, R.J. Science <u>268</u> (1995) 542.

42. P. Constantin, C. Foias, B. Nicolaenko, and R. Teman, <u>Integral and Inertial Manifolds for dissipative Partial Differential Equations</u> (Springer-Verlag, New York, 1980).

43. Per Bak, comment  made during a University  of Houston  colloquium (1994).

44. R. Olby,  <u>Origins of Mendelism</u>, 2nd Ed. (Univ.  of Chicago Pr., Chicago, 1985).

45. R. M. Young, <u>Darwin's Metaphor</u> (Cambridge Univ. Pr., Cambridge, 1985).

46. P. J. Bowler, <u>The Mendelian Revolution</u> (Johns Hopkins, Baltimore, 1989).

47. H. Morowitz, Complexity <u>1</u> (1995) 4.

48. E. P. Wigner, <u>Symmetries and Reflections</u> (Univ. Indiana Pr., Bloomington, 1967)

49. H. Friedrich, Physics World (Apr. 1992) pp. 32-36.

50. D. Lindley, <u>The End of Physics</u> (Basic Books, New York, 1993).

51. S. Weinberg, <u>Reductionism Redux</u> (New York Review of Books, 5 Oct. 1995), pp. 39-42.

**Footnotes**

a. "From a physicist's viewpoint, though, biology, history, and economics can be viewed as dynamical systems." [33]

b. By "dynamical system" in this context we mean deterministic and stochastic ordinary and partial differential equations, and also automata, including both cellular and noncellular automata, deterministic and stochastic.

c. Gregor Mendel was a "reductionist" in the Galilean tradition who got genetics right and thereby started the only mathematical science within biology. He was trained more as a physicist and mathematician than as a biologist. Mendel studied under Doppler, e.g., and even taught experimental physics [44]. Charles Darwin, in contrast, was a "holist" in the spirit of Aristotle. Darwin used nonmathematical ideas about "competition, natural selection and adaptation" that stemmed directly from Malthusian socio-economic policy [45] and therefore from Calvinism. Darwin his contemporaries (excepting Mendel) imagined an "integrated" picture of heredity that is completely wrong [46].

d. "Grand reductionism" consists essentially of the belief that the main job of physics is to propose a Lagrangian and then analyze its symmetries. "Petty reductionism" recognizes that writing down a Lagrangian is merely the beginning rather than the end of the job. To discover whether a Lagrangian or any other dynamical system has any significance it is first necessary to understand the stable, unstable, and complex motions (or their quantized analogues) that are generated by it, and then to see whether they have any significance for experiment and observation. Petty reductionism emphasizes that quarks, the standard model, and the rest of particle physics are completely irrelevant for understanding turbulence and other macroscopic natural phenomena (biology can not be "reduced" to quarks in any meaningful way). Awareness of ideas about complexity, which also belong to petty reductionism, suggests that it is unlikely that there can be a "final" theory of particle physics, and instead that more and more new structure ("surprises") would likely be found at shorter length scales if higher and higher energy experiments could be performed. The illusion of our ability to discover "final causes" goes back to Aristotle, and beyond, and does not have a scientific basis.

**Figure Caption**

Successive maxima $z_n$ of a numerically-computed time series $x_3(t)$ for the Lorenz model are plotted against each other (from reference [24]). The drawing of a single continuous curve through all of these points would violate the time-reversibility.